\documentclass[useAMS,usenatbib]{mn2e}

\usepackage{graphicx}

\def\cm{{\rm\thinspace cm}}

\def\erg{{\rm\thinspace erg}}
\def\g{{\rm\thinspace g}}
\def\Gyrs{{\rm\thinspace Gyrs}}
\def\km{{\rm\thinspace km}}
\def\K{{\rm\thinspace K}}
\def\kpc{{\rm\thinspace kpc}}

\def\Msun{\hbox{$\rm\thinspace M_{\odot}$}}

\def\yr{{\rm\thinspace yr}}

\def\s{{\rm\thinspace s}}

\def\kmps{\hbox{$\km\s^{-1}\,$}}

\def \etal {et~al.~}

\title{Cold stream stability during minor mergers}
\author[L. Wang \etal]
{\parbox{\textwidth}{Liang Wang$^{1,2}$\thanks{E-mail:
wangliang@pmo.ac.cn; wszhu@nju.edu.cn}, Weishan Zhu$^{3,4}$\footnotemark[1], Long-Long Feng$^{1}$, Andrea V. Macci{\`o}$^{2}$, Jiang Chang$^{1,2}$, Xi Kang$^{1}$}
\vspace{0.4cm}\\
$^{1}$ Purple Mountain Observatory, the Partner Group of Max-Planck-Institute for Astronomy, CAS, Nanjing, 210008, China\\
$^{2}$ Max Planck Instutut f\"ur Astronomie, K\"onigstuhl 17, 69117, Heidelberg, Germany\\
$^{3}$ School of Astronomy and Space Science, Nanjing University, Nanjing, 210093, China\\
$^{4}$ Key Laboratory of Modern Astronomy and Astrophysics of the Ministry of Education, Nanjing, 210093,
China}

\begin{document}

\date{Accepted . Received ; in original form}


\maketitle

\label{firstpage}

\begin{abstract}

We use high-resolution Eulerian simulations to study the stability
of cold gas flows in a galaxy size dark matter halo ($10^{12} \Msun$) at redshift $z=2$.
Our simulations show that a cold stream penetrating a hot
gaseous halo is stable against thermal convection and
Kelvin-Helmholtz instability. We then investigate  
the effect of a satellite orbiting the main halo in the plane of the stream.
The satellite is able to perturb the stream and to inhibit
cold gas accretion towards the center of the halo for 0.5 Gyr.
However, if the supply of cold gas at large distances
is kept constant, the cold stream is able to  re-establish itself 
after 0.3 Gyr. We conclude that cold streams
are very stable against a large variety of internal and external perturbations.

\end{abstract}


\section{Introduction}

How do galaxies get their gas is a key question to better understand their formation and 
evolution. In the conventional picture, the cold gas is accreted onto the Dark Matter halo, and
is shock heated to the virial temperature of the halo. The hot gas then loses pressure support due to radiative cooling, then
settles into a centrifugally supported disc and begins to form stars \citep{b9,b21}.
This picture has been supported by early simulations which have shown that a substantial fraction of cold gas 
was indeed shock  heated after accretion \citep{b3,b13}.

A number of theoretical and simulation-based works in the past decade, however, have found two
distinct modes of gas accretion, named as hot and cold accretion \citep{b4,b14,b7,b15,b8}. 
Cold gas flowing along dark matter filaments can be directly accreted onto the central galaxy without
being shock heated, cold streams are the main source of gas supply in galaxies especially in low-mass halos of $M_h \leq 10^{12} \Msun$ at high redshift \citep{b8}.

Simulations  also predict  that  these cold   streams,  being   optically  thick
Lyman-limit systems  with low  metallicities, could be  observable via
absorption   lines  against   background-objects \citep{bFumagalli2011}. 
\citet{b19} proposed that cold-flow streams will produce an
orbiting  circum-galactic  component of  cool  gas  with high  angular
momentum, which  might be observable as well.  However, the existence  of cold
streams has not  been confirmed by solid observational results.  Observations from \citet{b18}, showed
outflow rather than inflow of cold gas in the circum-galactic medium . 
\citet{b11}  argued that the covering  factor might be smaller than $10\% -20\%$ at $z=2$, 
so the observation of cold streams is very difficult. 
Very recently, several observations \citep{bBouche2013, b36} reported 
distinct signatures of strong metal poor absorbing systems in the vicinity of galaxies at $z \sim 2.3- 2.4$ 
by observing background quasars. Features of these systems are broadly 
consistent with the prediction of accreted cold gas in simulations.
However, to confirm the existence of cold streams,  information on the morphology of
the cold gas is needed.

If cold  streams  do appear  at some  epoch,  the velocity  shear
between them and hot gas will drive Kelvin-Helmholtz (K-H) instability.
When the contact surface  between cold  and  hot  gas is  not  aligned  with the  gravity
gradient,  Rayleigh-Taylor (R-T) instability  will  also  grow. 
In  addition,  thermal evaporation  will occur.  
Another source of instability for the cold flows may come from the very 
high merger rate of dark matter haloes predicted in the hierarchical scenario,
especially at high redshift \citep{b10}. 
All these possible sources of perturbation  naturally  raise the
question: how long could  the cold streams survive against
such instabilities?  

In this  Letter we make a first attempt  in trying to  determine the
stability  of cold  flows.  We employ  high resolution  hydrodynamical
Eulerian simulations combined with a  simplified set-up that allows us
to better understand the impact of all these different sources of
instability.  The  Letter is  organized as  follows:  We  present our
numerical  methods  and  cold   flow  model  in  \S2.   Hydrodynamical
simulation results  are shown in  \S3.  Discussion and  conclusion are
given in \S4.

 
\section[]{Initial parameters set-up and numerical simulations}

We assume that the cold streams are penetrating through a 
hot gaseous halo which is in hydrostatic equilibrium with the
gravitational potential of the hosting dark matter halo. 
The dark matter halo has a mass of $M_h=10^{12} M_{\odot}$ and
we assume a NFW profiles \citep{nfw97} with a
concentration parameter of 10 following \cite{m08}.
All the  cosmological parameters are set in agreement with 
the WMAP5 results \citep{b37}.

\subsection{Hot gaseous halo}
 
The hot halo gas is initially  set to be isothermal and in hydrostatic
equilibrium  under  the gravitational  potential  of  the dark  matter
halo.  The  self  gravity  of  gas is  not  included.  The central  
(stellar) galaxy component is not considered as we are mainly interested in 
the hydrodynamical evolution of  cold streams within the virial  
radius.
The temperature of the hot halo is set to  $T_{halo}  \sim  3.3  \times 10^6  \K$,
and we assume a  poly-tropic index  $\gamma=5/3$ polytropic. Finally we extend
the gas distribution to 1.5 times the virial radius of the dark
matter halo, (to what we call the boundary radius $R_b \equiv 1.5 R_h$), 
in order to reduce any possible boundary effect. The hot gas accounts for $17\%$ of the total mass within $R_h$.

\subsection{Cold stream}

The morphology, temperature, density and  velocity of cold streams are
set   according   to   recent   theoretical   and   simulation   works
\citep{b8,b16,b23}.  \citet{b23} performed a very broad statistical study of the 
cold streams velocity and temperature  as function of distance from the
halo center for dark  matter haloes with  $10^{11.5}<M_{h}<10^{12.5} \Msun$  at  $z=2$.  
According to their results we assume the velocity of cold streams at
$R_b$ to be  $v_{cold}(R_b) \sim 100 \kmps$.

The cold stream is firstly assumed to be a simple cone geometry, following simulations results \citep{b8, b15}.
The radius of the stream at the boundary $R_b$ is given by
\begin{equation}
R_{cold}(R_b) \sim R_b \times \sin(\frac{\theta}{2})
\end{equation}

\noindent
where  $\theta$ is  the  stream opening  angle and  could  be  derived 
from  the covering  factor   $\eta$  as  $\theta  \sim   2\pi  \eta  /x_{cold}$,
where $x_{cold}$  is the  number  of cold streams in the halo,  
we  adopt  the value  of $x_{cold}=3 $\citep{b6}.  
On the other hand the covering factor $\eta$ is more difficult to estimate.
\citet{b8} predicted that  gas at temperature $< 10^5
\K$  with column  densities $>  10^{20} \cm^{-2}$  should cover  
$\sim 25\%$ the area at radii from 20  to 100 kpc in haloes with $M_{h} \sim
10^{12} \Msun$  at $z  \sim 2.5$.   \citet{b11} measured  the covering
factor for  a Milky  Way type  progenitor Lyman-break galaxy in a cosmological 
simulation and  found a  smaller value,  which is  $\sim
10\%$ for the Lyman limit systems and $\sim 3\%$ for the damped Ly$\alpha$ absorbers within $R_h$.
We adopt a value of $20\%$  within $R_b$, in agreement with recent studies by \citet{bGoerdt2012,bFumagalli2013}. 
Therefore, $R_{cold}(R_b)\sim 15 \kpc$.

The density of cold gas $\rho_{cold}$ at $R_b$ can be determined by the accretion rate
$\dot{M}(R_b,t)$, $R_{cold}(R_b)$, the velocity and the mass ratio of
cold accretion mode as,
\begin{equation}\label{rhocold}
\rho_{cold}(R_b)=f_{cold}\frac{\dot{M}}{x_{cold}   v_{cold}(R_b)   \pi
  R^2_{cold}(R_b)}
\end{equation}
where, $f_{cold}$ is the cold accreted gas mass fraction. Using the so called
extended Press \& Schechter approach \citep{bLacey1994}
for structure formation, \citet{b8} derived  that the
growth rate of the baryonic mass at the virial radius should be
\begin{equation}
\dot{M} \simeq 6.6 \times M^{1.15}_{12}(1+z)^{2.25}f_{.165} ~~ \Msun \yr^{-1}
\end{equation}
where $M_{12} \equiv M_h/10^{12}\Msun$, and $f_{.165}$ is the baryonic
fraction in the  haloes in units of  the cosmic mean. 
For  a halo with $M_h=10^{12} \Msun$, it is about $\dot{M} \simeq 80 \Msun \yr^{-1}$ at
$z=2$. However, \cite{b23} found a  mean value  of $\sim  30 \Msun \yr^{-1}$  and $\sim  20 \Msun
\yr^{-1}$ for the total and cold baryonic mass accretion respectively.

For our selected cold accretion rate, $f_{cold}\times \dot{M}=20 \Msun \yr^{-1}$, and the number of cold streams, $x_{cold}=3$, the net accretion rate for one single cold stream is $7 \Msun \yr^{-1}$. In our simulation we only model one such cold stream. The density of the cold stream at the boundary, $R_{b}$ is $\sim 3.0 \times 10^{-26} \g \cm^{-3}$ obtained from Eq.2. The temparature of cold gas is set to $T_{cold}(R_b)=3.3 \times 10^{4} \K$, which is almost in thermal pressure quilibrium. In Fig.1 we give a schematic of the initial condition, and the simulation parameters are listed in table~\ref{table}.

\subsection{Orbiting satellite}

The satellite is described by a single, mass varying particle with
an initial mass of  $3\times  10^{10}$  $\Msun$, leading to 
a  merger mass  ratio  of 1/33.  The orbit of the satellite is predetermined and extracted from a similar high resolution N-body simulation in \citet{b5}. The initial position of the satellite is at $R_b$, locating at the opposite side of the cold accretion with respect to the halo center. The initial radial and tangential velocity of the satellite are respectively set to $0.9$ and $0.6 $ times of the virial 
velocity of the main halo, $v_{vir}=160 \kmps$.The mass evolution
of the satellite is also extracted, giving $ 1.4 \times 10^{10} \Msun$ at 2 Gyr.

\begin{figure}
\centering
 \includegraphics[width=0.32\textwidth]{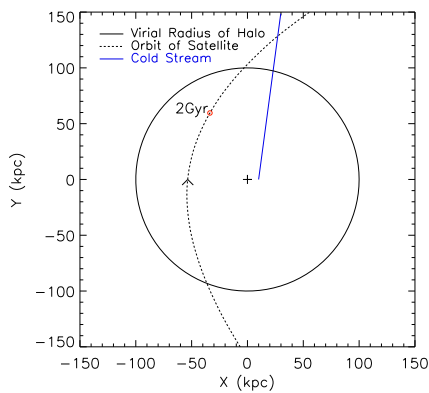}
 \caption{The sketch map of our simulation. The solid black circle is
   the virial  radius of dark matter  halo, the blue sold  line is the
   axis of injected cold  stream at $R_b$, the dashed curve is the  orbit of the satellite and
  the  red circle mark the position of the  satellite at 2 Gyrs.}
\label{figure1}
\end{figure}

Finally we assume the orbit plane of the
satellite is co-planar with the  cold stream plane. Such a choice is motivated
by results from N-body simulations \citep{b38}, and by 
recent observation of a possible ``cosmic alignment'' of galactic satellites
in M31 \citep{b12}. 
A sketch of our initial configuration and the satellite orbit are presented in
Figure \ref{figure1}.

\begin{table}
 \centering
 \begin{minipage}{140mm}
  \caption{Parameters used in the hydrodynamic simulation}
  \label{table}
  \begin{tabular}{@{}ll@{}} 
  \hline halo  concentration (c)& 10  \\ virial radius ($R_h$)  & $100
  \kpc$ \\  virial mass ($M_h$) &  $10^{12} \Msun$ \\ hot  gas central
  density ($\rho_{hot}$)  & $3.0 \times  10^{-24} \g \cm^{-3}$  \\ hot
  gas  temperature ($T_{hot}$)  & $3.3  \times  10^6 \K$  \\ boundary radius $R_b$ & $150 \kpc$ \\ cold  gas
  density at $R_b$ ($\rho_{cold}(R_b)$)  &  $3.0  \times 10^{-26}  \g  \cm^{-3}$ \\  cold
  gas velocity at $R_b$ ($v_{cold}(R_b)$) &  $100 \kmps$  \\ cold  gas temperature at $R_b$
  ($T_{cold}(R_b)$) & $4.0 \times 10^4 \K$  \\ cold gas pressure at $R_b$ & $ 10^{-15}
  \erg \cm^{-3} $ \\ \hline
\end{tabular}
\end{minipage}
\end{table}

\subsection{Numerical methodology}

To track the evolution of cold streams,  we run
hydrodynamical simulations with a fixed grid code using the positive preserving WENO scheme
\citep{b24},  which can effectively keep positive density and  pressure  in
hypersonic regions, and can well capture the motions of the baryonic gas in
cosmological  simulations \citep{b25}. Simulations are performed  in a
cubic box of 300 kpc,  with a
number of grid cells $512^3$,  i.e, a  resolution of  $0.59$ $\kpc$. Analytical gravitational potentials
based on NFW profiles are given at each grid cell initially. 
We impose hydrostatic equilibrium for the boundary condition except for the stream.
Density and velocity in the ghost grid cells outside the boundary of the simulation box are assumed to  be zero-gradient along  the radial
direction and the pressure is given by a first-order differentiation of the
hydrostatic equilibrium equation.

Radiative cooling  is followed by  using the cooling function  given in
\citet{b20}. Recent  simulation works \citep{b23,b16} showed  that the
metallicity  abundance ranges from   $\sim  10^{-1}$   to  $10^{-3}$
$Z_{\odot}$  for the  cold gas.  We  adopt the  value $\sim  10^{-1.5}
Z_{\odot}$,  and further assume the hot gas has the  same value
for the sake of simplicity. Thermal conduction is  also included with
the unsaturated regime assumption, and the  thermal conductivity is set to
$\kappa(T)=\kappa_0 T^{5/2}$  \citep{b17}, with $\kappa_0  \simeq 1.84
\times 10^{-5}  (\ln \Lambda)^{-1}  \erg \s^{-1}  \cm^{-1} \K^{-7/2}$,
where $\ln  \Lambda$ is  the Coulomb logarithm,  which is  only weakly
dependent on $n_e$ and $T$ ($\ln \Lambda = 30$ in our simulation).

From our model set up, the cold gas is injected into the simulation box within an aperture radius of $15$ kpc  at $R_b$ with a constant rate of $7 \Msun \yr^{-1}$, with the density, velocity and temperature given in Table.1. The constant inflow rate lasts for the whole simulation run, i.e., 4 Gys. It is found that a stable cold stream towards the halo central region is soon established under the gravity of the host halo, with time less than 1.5 Gyr (top left panel of Fig.2). Note that for the cold stream we assume an impact parameter $b\sim 30 \kpc$ respect to the halo center, as suggested by cosmological simulations \citep{b8,b15}.


We perform two sets of simulations, with and without the inclusion of the 
merging satellite, which are referred to
as \textit{col-mer-512}  and \textit{col-ref-512}  respectively.  
The latter simulation is  intended to  track the  effect of  thermal conduction  and K-H
instability, while the former is used to study  the impact of an orbiting 
satellite on the stability of cold flows.


\section{Simulation Results}

\begin{figure*}
  \includegraphics[width=0.78\textwidth]{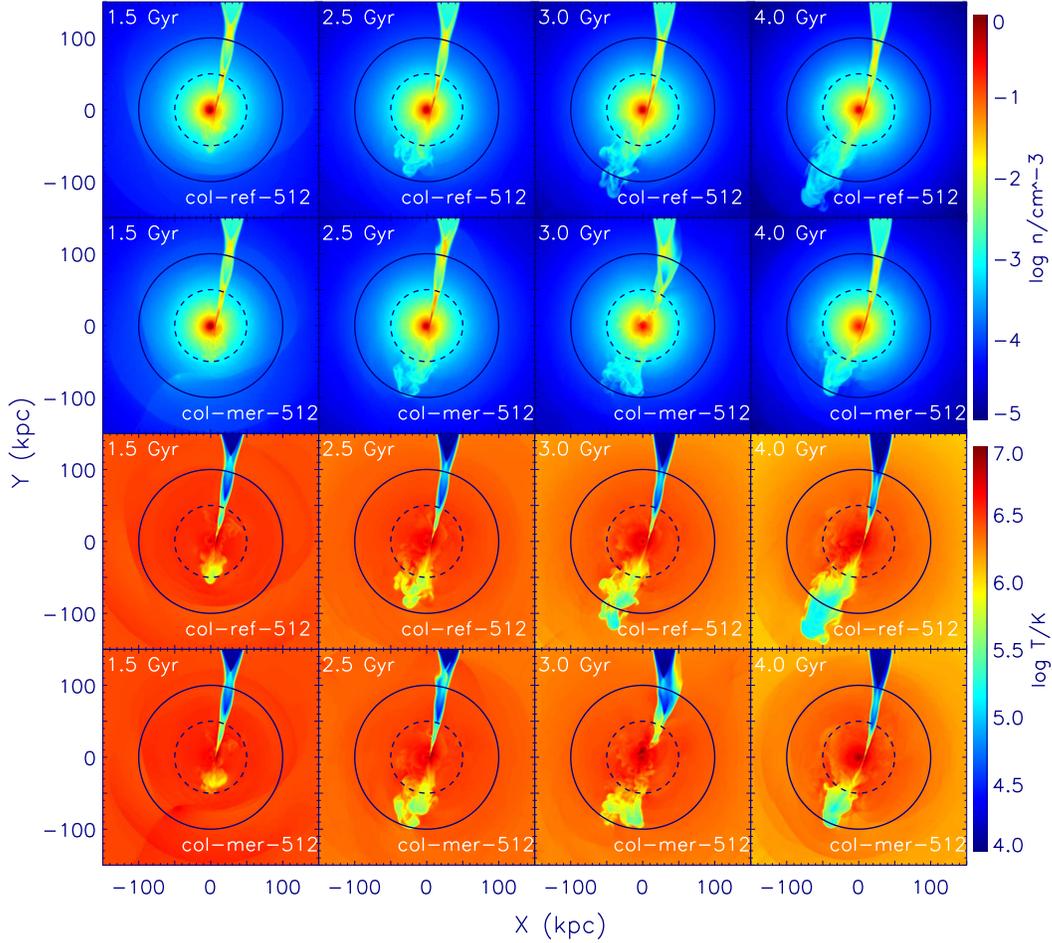}
  \caption[figure2]{The  top  two rows  are  density  slices of  the
    \textit{col-ref-512}    and   \textit{col-mer-512}    simulations,
    (1st  and  2nd  row   respectively).  Each  column
    represents 1.5,  2.5, 3.0 and  4.0 $\Gyrs$. The bottom  two rows
    are temperature slices. The solid/dashed line indicates the virial radius $R_h$/0.5$R_h$. At $t=1.5$ Gyr the radius of the stream at $R_h$ and $0.5 R_h$ is around 7.5 and 5 kpc respectively.} 
  \label{figure2}
\end{figure*}

Fig.~\ref{figure2} presents the slices of density and temperature
distribution  of   gas  at  1.5,  2.5,   3.0  and  4.0  Gyrs   in  the
simulations without/with the satellite in the upper/lower panels, respectively. 
The evolution before 1.5 Gyrs, during the formation of the stream, is practically the same
in the two simulations and it is not shown.

In the simulation without satellite, only subtle
changes in morphology, density and temperature
happen to the stream after its formation.  
Thermal conduction  results a thin transition  layer between the two
gas phases and evidence for K-H  instability is  hard  to find. The cold gas could flow into the central region ($< 0.3 R_h$) with a velocity as high as $400 \kmps$. 
After passing through the inner region, a  small fraction of gas flows out of
$0.5 R_h$. In a full cosmological simulation this gas will be retained by the central galaxy
and used for star formation.

In  the  \textit{col-mer-512} simulation,  obvious  disturbances  have  been
triggered by the  satellite at 2.5 and  3.0 Gyrs. There is practically no 
flow of cold gas inside $0.5 R_h$ at 3.0 Gyr. Interestingly enough the
cold stream is re-established after 4.0 Gyr,
when the satellite is far away from the stream. 

\begin{figure}
\centering
  \includegraphics[width=0.38\textwidth]{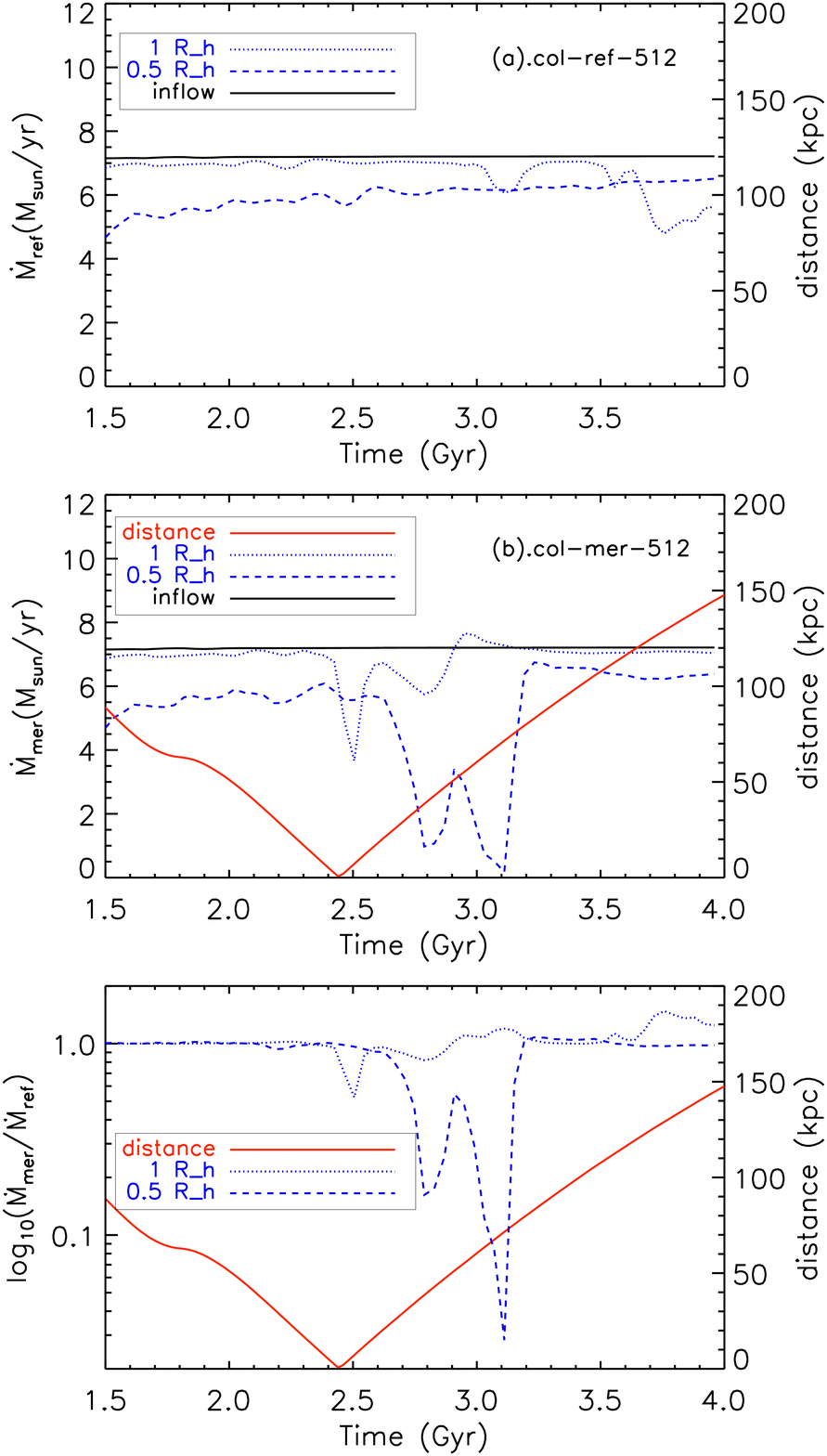}
  \caption[figure3]{The evolution of the cold gas accretion rate. Top/Middle  panel: for the simulation without/with a merging satellite with a 1:33 mass ratio. 
    Bottom  panel: the ratio of gas accretion between
    the two simulation runs.  The black line indicates the cold gas inflow rate at the boundary of the simulation box ($1.5 R_h$). In the  middle and bottom
    panels, the  red line indicate  the distance between  the satellite
    and the cold stream.}
  \label{figure3}
\end{figure}

In order to quantify the effect of  satellite on the cold stream,
we present in Fig.~\ref{figure3}  the evolution of the net accretion rate of cold gas through
a spherical shell at $R_h$ and at $0.5 R_h$. Gas with temperature below $3\times 10^5$ K is defined as cold.   
The \textit{col-ref-512} run shows an almost constant accretion rate at
both radius. This implies that thermal  conduction and gravitational heating are able to
convert only a small fraction  of the incoming cold gas to the hot phase.

The situation is very different for the \textit{col-mer-512} run.  
The accretion rate at both radius shows sharp drops delayed by a couple of hundred 
Myr with respect to the crossing of the stream by the satellite. The accretion rate at at 0.5 $R_h$  
is practically suppressed for a period  of $\sim  0.5$
Gyrs.  This very reduced cold accretion is clearly connected with the close
passage of the satellite at $t=2.5$ Gyrs, marked by the red line in the figure.

Despite the very destructive effect of the satellite on the stream,
the cold flow is able to re-build itself quite rapidly in less
than 0.3 Gyrs. This shows that in the presence of a continuous
accretion of cold gas from cosmological filaments, even if disturbed
by orbiting satellites, cold flows are able to regenerate
themselves quite rapidly.

\section{Discussion and Conclusion}

In this letter, inspired by  recent simulation works and observations,
we have studied the stability  of cold streams penetrating through hot
gaseous halos in dark matter  halos with mass $M=10^{12} M_{\odot}$ at
$z=2$  by  hydrodynamic simulations  in  a  simplified scenario.

Without external perturbations (e.g. satellites), 
the cold gas is able to flow into the inner region of the central galaxy. 
The K-H instability
and thermal conduction are not able to inhibit the cold gas accretion.
K-H instability might have
been damped by the supersonic motion of the gas 
due to the low cooling rate \citep{bVietri1997}.

When we consider the disturbing action of a satellite (with a mass ratio
of 1:33) merging with the central halo,
the  accretion rate of cold gas at  half of the virial radius experiences a
dramatic drop  for  $\sim$ 0.5 Gyrs. In this period
the supply of cold gas into the central region 
decreases by more than $\sim 70\%$.

Cosmological simulations have shown that the merger rate at $z=2$ 
for our satellite/host mass combination is about two mergers
per Gyr\citep{b10}. However, the cross section between 
 isotropically distributed stream and satellites can be crudely estimated by
\begin{equation}
\frac{V_{cs}}{V_{halo}} = \left(R_{cold}\right)^2 2\pi
     \frac{R_h}{3} / \frac{4\pi}{3} R_h^3 \sim 1\%
\end{equation}
where $V_{cs}$ and $V_{halo}$ are the volume of cold stream and halo respectively.
Therefore, the realistic distribution of cold stream and satellite would play an important role.

Even in the case of a direct interaction between the satellite and 
the stream, the flow of cold gas towards the halo center is able to re-establish
itself in less than 0.3 Gyrs, under the assumption of a continuous 
flow of cold gas from cosmic filaments.

What we present in this Letter is a pilot study for the stability 
of Cold Flows. We plan to improve and extend it in forthcoming
works by increasing the resolution of the simulation, considering
different satellite orbits and impact parameters, adding the
self gravity of the gas and comparing
different hydrodynamical approaches (see \citet{bNelson_2013}
for the effect of the hydro solver on the cold flow stability
in cosmological simulations).
Meanwhile, we expect that further observations may reveal
more information about both cold and hot gas in the halo, and more
details on galaxy formation and growth.

\section*{Acknowledgments}
We acknowledge useful discussions with Greg Stinson and Salvo Cielo, We thank Kester Smith for revising the text. 
The simulations are performed at the
Hyper Performance  Computing Center of Nanjing  University. 
LW and WSZ thank  the MPIA for its hospitality during the preparation of this work. 
WSZ thanks the support  from the  China Postdoctoral Science Foundation  and the
Fundamental  Research Funds  for the  Central Universities.  WSZ, LLF and XK
are supported  by the  National  Natural Science  Foundation of  China
under grant 11203012, 11273060, 91230115, 11333008, 11073055. LW, AVM,JC and XK acknowledge support from the MPG-CAS through the partnership program between 
the MPIA group lead by A. Macci\`o and the PMO group lead by X. Kang.


\begin{thebibliography}{99}

\bibitem[\protect\citeauthoryear{Benson et al.}{2001}]{b3} Benson A.J., Frenk C.S., Baugh C.M., Cole S., Lacey C.G., 2001, MNRAS, 327, 1041
\bibitem[\protect\citeauthoryear{Birnboim \& Dekel}{2003}]{b4} Birnbiom Y., Dekel A., 2003, MNRAS, 345, 349
\bibitem[\protect\citeauthoryear{Bouch{\`e} et al.}{2013}]{bBouche2013} Bouch{\`e} N., Murphy M.T. et al., 2013, Science, 341, 50
\bibitem[\protect\citeauthoryear{Chang et al.}{2013}]{b5} Chang J., Macci{\`o} A.V., Kang X., 2013, MNRAS, 431, 3533
\bibitem[\protect\citeauthoryear{Crighton et al.}{2013}]{b36} Crighton N., Hennawi J., Prochaska J., 2013, ApJL, 776, L18
\bibitem[\protect\citeauthoryear{Danovich et al.}{2012}]{b6} Danovich M., Dekel A., Hahn O., Teyssier R., 2012, MNRAS, 422, 1732
\bibitem[\protect\citeauthoryear{Dekel \& Birnboim}{2006}]{b7} Dekel A., Birnbiom Y., 2006, MNRAS, 368, 2
\bibitem[\protect\citeauthoryear{Dekel et al.}{2009}]{b8} Dekel A., Birnbiom Y.,Engel G. et al., 2009, Nature, 457, 451
\bibitem[\protect\citeauthoryear{Fall \& Efstathiou}{1980}]{b9} Fall S.M., Efstathiou G., 1980, MNRAS, 193, 189
\bibitem[\protect\citeauthoryear{Fakhouri et al.}{2010}]{b10} Fakhouri O., Ma, C-P., Boylan-Kolchin M., 2010, MNRAS, 406, 12

\bibitem[\protect\citeauthoryear{Faucher-Gigu{\`e}re \& Kere{\v s}}{2011}]{b11} Faucher-Gifu{\`e}re C.A., Kere{\v s} D., 2011, MNRAS, 412, L118
\bibitem[\protect\citeauthoryear{Fumagalli et al.}{2011}]{bFumagalli2011} Fumagalli M., Prochaska J., Kasen D., Dekel A., Ceverino D., Primack J. R., 2011, MNRAS, 418, 1796
\bibitem[\protect\citeauthoryear{Fumagalli et al.}{2013}]{bFumagalli2013} Fumagalli M., {\`O}Meara J. M., Prochaska J. X., Worseck G., 2013, ApJ, 775, 78
\bibitem[\protect\citeauthoryear{Goerdt et al.}{2012}]{bGoerdt2012} Goerdt T., Dekel A., Sternberg A., Gnat O., Ceverino D., 2012, MNRAS, 424, 2292
\bibitem[\protect\citeauthoryear{Helly et al.}{2003}]{b13} Helly J.C., Cole S., Frenk C.S., Baugh C.M., Benson A., Lacey C., Pearce F.R., 2003, MNRAS, 338, 913
\bibitem[\protect\citeauthoryear{Ibata et al.}{2013}]{b12} Ibata R.A., Lewis G.F. et al., 2013, Nature, 2013, 493, 62
\bibitem[\protect\citeauthoryear{Kere{\v s} et al.}{2005}]{b14} Kere{\v s} D., Katz N., Weinberg D.H., Dave R., 2005, MNRAS, 363, 2
\bibitem[\protect\citeauthoryear{Kere{\v s} \& Hernquist}{2009}]{b15} Kere{\v s} D., Hernquist L., 2009, ApJ, 700, L1

\bibitem[\protect\citeauthoryear{Komatsu et al.}{2009}]{b37} Komatsu, E., Dunkley, J., Nolta, M. R., et al. 2009, ApJS, 180, 330

\bibitem[\protect\citeauthoryear{Lacey \& Cole}{1994}]{bLacey1994} Lacey C., Cole S., 1994, MNRAS, 271, 676
\bibitem[\protect\citeauthoryear{Macci{\`o} et al.}{2008}]{m08} Macci{\`o}, A.~V., 
Dutton, A.~A., \& van den Bosch, F.~C.\ 2008, MNRAS, 391, 1940 

\bibitem[\protect\citeauthoryear{Navarro et al.} {1997}]{nfw97} Navarro, J.~F., Frenk, 
C.~S., \& White, S.~D.~M. 1997, ApJ, 490, 493 

\bibitem[\protect\citeauthoryear{Nelson et al.}{2013}]{bNelson_2013}  Nelson, D., et al. 2013, MNRAS, 429,3353
\bibitem[\protect\citeauthoryear{Ocvirk et al.}{2008}]{b16} Ocvirk P., Pichon C., Teyssier R., 2008, MNRAS, 390, 1326
\bibitem[\protect\citeauthoryear{Spitzer}{1962}]{b17} Spitzer L., 1962, Physics of Fully Ionized Gases (New York: Interscience)
\bibitem[\protect\citeauthoryear{Steidel et al.}{2010}]{b18} Steidel C., Erb D. Shapley A. et al., 2009, ApJ, 700, L1
\bibitem[\protect\citeauthoryear{Stewart et al.}{2010}]{b19} Stewart K.R., Kaufmann T., Bullock J.S., Barton E.J., et al., 2011, ApJ, 738, 39
\bibitem[\protect\citeauthoryear{Sutherland \& Dopita}{1993}]{b20} Sutherland R.S., Dopita M.A., 1993, ApJS, 88, 253
\bibitem[\protect\citeauthoryear{Vietri et al.}{1997}]{bVietri1997} Vietri M., Ferrara A., Miniati F., 1997, ApJ, 483, 262
\bibitem[\protect\citeauthoryear{White \& Frenk}{1991}]{b21} White Simon D. M., Frenk Carlos S., 1991, ApJ, 52, 379
\bibitem[\protect\citeauthoryear{van de Voort \& Schaye}{2012}]{b23} van de Voort F., Schaye J., 2012, MNRAS, 423, 2991
\bibitem[\protect\citeauthoryear{Zentner, Kravtsov, \& Gnedin et al.}{2005}]{b38}Zentner A. R., Kravtsov A. V., Gnedin O. Y., Klypin A. A., 2005, ApJ, 629,
219
\bibitem[\protect\citeauthoryear{Zhang \& Shu}{2010}]{b24} Zhang X., Shu C.W., 2010, Journal of Computational Physics, 229, 8918
\bibitem[\protect\citeauthoryear{Zhu, Feng, \& Xia et al}{2013}]{b25}  Zhu, W.S., Feng, L.L., Xia, Y.H., Shu, C.-W., Gu, Q.S., \& Fang, L. Z., 2013, ApJ, 777, 48 
\end{thebibliography}
\end{document}